\documentclass[journal,comsoc]{IEEEtran}
\usepackage[T1]{fontenc}
\usepackage[linesnumbered,ruled]{algorithm2e}
\usepackage{stfloats,color}
\usepackage{amsfonts}
\usepackage{amssymb}
\usepackage[cmex10]{amsmath}
\usepackage{graphicx}
\usepackage{setspace}
\usepackage{cite}
\usepackage{array}
\usepackage{mdwmath}
\usepackage{mdwtab}
\usepackage{times}
\usepackage{epsfig}
\usepackage{latexsym}
\usepackage{epstopdf}
\usepackage{verbatim}
\usepackage{units}
\usepackage{amsthm}
\usepackage{mdwlist}
\usepackage{placeins}
\usepackage{afterpage}
\usepackage{dsfont}
\usepackage{soul}
\usepackage{subcaption}

\ifCLASSINFOpdf
\else
\fi
\usepackage{amsmath}
\usepackage[cmintegrals]{newtxmath}
\newcommand{\defeq}{\ensuremath{\triangleq}}

\begin{document}
%
\title{ \LARGE Mobility and Popularity-Aware Coded Small-Cell Caching}
%
%
\author{Emre Ozfatura and
       Deniz G{\"u}nd{\"u}z
\thanks{Emre Ozfatura and Deniz G{\"u}nd{\"u}z are with the Information Processing and Communications Lab, Department of Electrical and Electronic Engineering, Imperial College London, London SW7 2AZ, UK. }
\thanks{This work was supported by EC H2020-MSCA-ITN-2015 project SCAVENGE under grant number 675891, and by the European Research Council project BEACON under grant number 677854.}}

%
%

\markboth{}%
{Shell \MakeLowercase{\textit{et al.}}: Bare Demo of IEEEtran.cls for IEEE Communications Society Journals}
%



\maketitle
\begin{abstract}
In heterogeneous cellular networks with caching capability, due to mobility of users and storage constraints of small-cell base stations (SBSs), users may not be able to download all of their requested content from the SBSs within the delay deadline of the content. In that case, the users are directed to the macro-cell base station (MBS) in order to satisfy the service quality requirement. Coded caching is exploited here to minimize the amount of data downloaded from the MBS taking into account the mobility of the users as well as the popularity of the contents. An optimal distributed caching policy is presented when the delay deadline is below a certain threshold, and a distributed greedy caching policy is proposed when the delay deadline is relaxed. 
\end{abstract}

%
\IEEEpeerreviewmaketitle
\vspace*{-3mm}
\section{Introduction}
%
%
%
%
\IEEEPARstart{C}{aching} popular video contents at the network edge, closer to the end users, is a promising method to cope with increasing video traffic. In heterogeneous cellular networks, popular video files are stored at SBS caches to reduce latency as well as backhaul traffic \cite{hetcont1,hetcont2,femto}. The goal in these works is to maximize the hit rate, or equivalently, to minimize the amount of data downloaded from MBSs under given SBS cache capacity constraints. A common assumption is that a user stays connected to the same set of SBSs during the whole duration of video download. However, in ultra dense networks, where there is a large number of operating SBSs with limited coverage areas, it is indispensable to take user mobility into consideration to meet the prescribed quality of service (QoS) requirements \cite{mobilityint}.

Our aim in this paper is to provide a caching policy, which, for given video popularity profile and user mobility patterns, minimizes the average amount of data downloaded from the MBS. We adopt the {\em delayed offloading} scheme in heterogeneous networks; that is, each video content request has a deadline, and if the mobile user cannot download all the fragments of the content from the SBSs it has connected to by the deadline, the remainder of the request is satisfied by the MBS \cite{offload}. We consider maximum distance separable (MDS) coding for storing the video contents in the SBS caches. To the best of our knowledge, \cite{mobpathcache1} is the only work that considers mobility-aware caching with delayed offloading. The goal in \cite{mobpathcache1} is to minimize the probability of a request being served by the MBS. In this work, as in \cite{MBScache}, we assume that the MBS serves the remaining fragments as MDS coded packets at a higher cost. Hence, unlike \cite{mobpathcache1}, our goal is to minimize the amount of data downloaded from the MBS. We show that, if the request deadline is below a certain threshold there is an optimal distributed solution. If the request deadline does not meet this condition, we introduce a sub-optimal greedy caching policy. 
\vspace*{-3mm}  
\section{System Model and Problem Formulation}
Consider a heterogeneous network that consists of one MBS and $N$ SBSs, ${\mathrm{SBS_{1}},\ldots,\mathrm{SBS_{N}}}$, with disjoint coverage areas. A mobile user (MU) is served by only one SBS at any particular time. We consider a video library $\mathbb{V}=\left\{v_{1},\ldots,v_{K}\right\}$ consisting of $K$ distinct video files, each of length $B$ bits, indexed according to their popularity profile, i.e., file $v_{k}$ is the $k$th most popular file with request probability $p_{k}$. Further, we assume that video files are encoded by rateless MDS coding \cite{femto}, so that a video file can be retrieved when $B$ parity bits are collected in any order, from any SBSs.

\indent We consider equal-length time slots, whose duration corresponds to the minimum time that a MU remains in the coverage area of a SBS. Therefore, although a MU cannot connect to more than one SBS during one time slot, it may stay connected to the same SBS over several consecutive time slots. We assume that $\mathrm{SBS_{n}}$ is capable of transmitting $R_{n}$ bits within a time slot to a MU within its coverage area, and it has a storage capacity of $C_{n}$ bits.  Due to the QoS requirement, a video file must be downloaded within $T$ time slots once it is requested. Thus, if a MU is not able to collect $B$ bits from SBSs in $T$ slots, the remaining bits are provided by the MBS at a higher cost. 

\indent We define the {\em mobility path} of a user as the sequence of small-cells visited over $T$ time slots after requesting a video file. For instance, for $T=5$, ${\mathrm{SBS_{1}},\mathrm{SBS_{2}},\mathrm{SBS_{2}},\mathrm{SBS_{3}},\mathrm{SBS_{4}}}$ is a possible mobility path. We remark that the MU may remain connected to the same SBS more than one time slot, and can connect to at most $T$ different SBSs. Hence, there is a finite number of distinct mobility paths, denoted by $M$. We  denote the $m$th mobility path by $I_{m}$, and its  realization probability by $q_{m}$. Realization probabilities can be obtained from empirical observations, or via modeling mobility paths as random walks on a Markov chain.\\   
\indent Our aim is to minimize the expected amount of data downloaded from the MBS for given SBS storage capacities $\mathbf{C}\defeq\left\{C_{n}\right\}^{N}_{n=1}$, data transmission rates $\mathbf{R}\defeq\left\{R_{n}\right\}^{N}_{n=1}$, video popularity profiles $\mathbf{P}\defeq\left\{p_{k}\right\}^{K}_{k=1}$, and mobility paths with realization probabilities $\mathbf{I}_{T}\defeq\left\{(I_{m},q_{m})\right\}^{M}_{m=1}$. Let $\mathbf{X}_{T}=\left\{x_{n,k}\right\}^{N,K}_{n=k=1}$ denote the caching policy over $T$ time slots, where $x_{n,k}$ indicates the number of parity bits for file $v_{k}$ stored in $\mathrm{SBS_{n}}$. A caching policy is {\em feasible} if $\sum^{K}_{k=1}x_{n,k}\leq C_{n}$, $\forall n$. The average amount of data downloaded from the MBS is denoted by $d_{av}(\mathbf{X}_{T},\mathbf{I}_{T})$, or simply by $d_{av}$.  \\ 
\indent Let $d_{k,m}$ denote the amount of coded data downloaded from the MBS for video $v_{k}$ following mobility path $I_{m}$. For given $\mathbf{C},\mathbf{R},\mathbf{P},\mathbf{I}_{T}$, and caching policy $\mathbf{X}_{T}$, we have
\begin{equation}\label{dkm}
d_{k,m}=\max\left\{B-\left(\sum^{N}_{n=1}\min \left\{x_{n,k},R_{n}S_{m,n}\right\}\right),0\right\},
\end{equation}
where $S_{m,n}$ denotes the total number of time slots the MU connected to $\mathrm{SBS_{n}}$ in mobility path $I_{m}$. Then, taking the average over all mobility paths and video files, $d_{av}$ can be written in terms of $d_{k,m}$ as $d_{av}=\sum^{M}_{m=1}\sum^{K}_{k=1}q_{m}p_{k}d_{k,m}$. Our goal is to find the optimal feasible caching policy $X^{\star}_{T}$ that minimizes $d_{av}$, formulated as follows:
\begin{small}
\begin{align}
   \text{\bf P1:} \;\;\;\min_{\mathbf{X}_{T}}&
   \begin{aligned}[t]
      d_{av}  \notag
   \end{aligned}\\
   \text{subject to: } &\sum^{K}_{k=1}x_{n,k}\leq C_{n}, \text{ }\forall n.\\
	& x_{n,k}\geq 0, \text{ }\forall n,k.
\end{align}
\end{small}
In order to avoid the $\max$ operator in Eqn. (\ref{dkm}), problem \textbf{P1} is reformulated as follows. We treat $d_{k,m}$ as decision variables, and add the following constraint:
\begin{equation}
d_{k,m}\geq\max\left\{B-\left(\sum^{N}_{n=1}\min \left\{x_{n,k},R_{n}S_{m,n}\right\}\right),0\right\}.\label{optconst}
\end{equation}
We note that, for a given feasible caching policy the objective function is monotonically increasing with $d_{k,m}$; hence, for the optimal solution, constraint (\ref{optconst}) must be satisfied with equality. The equivalent optimization problem is obtained as follows.
\begin{small}
\begin{align}
   \text{\bf P2:} \;\;\;\min_{\mathbf{X}_{T},\mathbf{D}_{T}}&
   \begin{aligned}[t]
       d_{av}  \notag
   \end{aligned} \\
  \text{subject to: }  &\sum^{K}_{k=1}x_{n,k}\leq C_{n}, \text{ }\forall n,\\
   &  \sum^{N}_{n=1}\min\left\{x_{n,k},R_{n}S_{m,n}\right\}+d_{k,m}\geq B, \text{ }\forall k,m,\label{optconst1}\\
	 &   d_{k,m} \geq 0, \text{ }\forall k,m,\label{optconst2}
\end{align}
\end{small}where $\mathbf{D}_{T}\defeq\left\{d_{k,m}\right\}^{K,M}_{k=m=1}$. Notice that constraints (\ref{optconst1}) and  (\ref{optconst2}) together imply  constraint (\ref{optconst}). Recall that a MU cannot connect to more than $T$ different SBSs. Hence, for each $(k,m)$ pair, constraint (\ref{optconst1}) can be replaced by at most $2^{T}$ linear constraints. To clarify, let $T=4$ and  $N=6$, and consider the mobility path $I_{m}=\left\{\mathrm{SBS_{1}},\mathrm{SBS_{1}},\mathrm{SBS_{2}},\mathrm{SBS_{2}}\right\}$. For this specific mobility path and  file $v_{k}$, (\ref{optconst1}) can be written as
\begin{equation}
\min\left\{x_{1,k},2R_{1}\right\}+\min\left\{x_{2,k},2R_{2}\right\}+d_{k,m}\geq B \label{example}.
\end{equation}
Equivalently, (\ref{example}) can be replaced by the following set of linear constraints:
\begin{align}
x_{1,k}+x_{2,k}+d_{k,m}&\geq B,\\
x_{1,k}+2R_{2}+d_{k,m}&\geq B,\\
2R_{1}+x_{2,k}+d_{k,m}&\geq B,\\
2R_{1}+2R_{2}+d_{k,m}&\geq B.
\end{align}
Consequently, the initial optimization problem \textbf{P1} can be cast into a linear optimization problem. However, the number of constraints are exponential in  time constraint $T$. In the next subsection we show that, under a certain assumption on $T$,  \textbf{P1} can be solved in a distributed manner.
\vspace*{-3mm}
\section{Distributed Solution}
In this section, we consider the case $T\leq T_{min} \defeq \frac{B}{R_{max}}$, where $R_{max}$ is the maximum data rate across all the cells, i.e., $R_{max} \defeq\max\left\{R_{1},\ldots,R_{N}\right\}$. This special case is also instrumental in highlighting the distinction between our problem formulation and that of \cite{mobpathcache1}, whose goal is to minimize the probability of downloading any data from the MBS. We note that, with the formulation of \cite{mobpathcache1}, when $T<T_{min}$ all caching policies are equivalent since it is not possible to collect $B$ bits in $T$ slots. While \cite{mobpathcache1} ignores the mobility paths when $T<T_{min}$, each caching policy will induce a different $d_{av}$. Hence, an optimal caching policy $\mathbf{X}_{T}$ in \cite{mobpathcache1} may lead to a suboptimal $d_{av}$. Instead, we present the optimal caching algorithm that minimizes $d_{av}$ when $T\leq T_{min}$. We also propose a greedy caching policy, for $T>T_{min}$. 
\vspace*{-3mm}
\subsection{Optimal Distributed Solution}
When $T\leq T_{min}$, (1) simplifies to 
\begin{equation}
d_{k,m}=B-\sum^{N}_{n=1}\min \left\{x_{n,k},R_{n}S_{m,n}\right\}.
\end{equation}
Then, our objective $d_{av}$ can be rewritten as:
\begin{equation}\label{daveqn}
d_{av}=B-\underbrace{\sum^{M}_{m=1}\sum^{K}_{k=1}\sum^{N}_{n=1}q_{m}p_{k}\min \left\{x_{n,k},R_{n}S_{m,n}\right\}}_{\defeq \tilde{d}_{av}}.
\end{equation}
Note that minimizing $d_{av}$ is equivalent to maximizing $\tilde{d}_{av}$, which denotes the average amount of data downloaded from the SBSs. We change the order of the summations in (\ref{daveqn}):
\begin{equation}
\tilde{d}_{av}=\sum^{N}_{n=1}\underbrace{\sum^{K}_{k=1}p_{k}\sum^{M}_{m=1}q_{m}\min \left\{x_{n,k},R_{n}S_{m,n}\right\}}_{\defeq \tilde{d}_{av,n}},
\end{equation}
we observe that the optimal caching policy can be obtained via maximizing $\tilde{d}_{av,n}$, defined above, for each $\mathrm{SBS_{n}}$ separately. Let $\mathbf{X}^{n}_{T}$ denote the caching policy for $\mathrm{SBS_{n}}$. For $\mathrm{SBS_{n}}$, we have the following optimization problem:
\begin{small}
\begin{align}
   \text{\bf P3:} \;\;\;\max_{\mathbf{X}^{n}_{T}}&
   \begin{aligned}[t]
      \tilde{d}_{av,n}  \notag
   \end{aligned}\\
   \text{subject to: } &\sum^{K}_{k=1}x_{n,k}\leq C_{n}. \label{const} \\
	 & x_{n,k}\geq 0, \text{ }\forall k.
\end{align}
\end{small} 
If we group the mobility paths according to the time spent in cell $\mathrm{SBS_{n}}$, $\tilde{d}_{av,n}$ can be written as
\begin{equation}
\tilde{d}_{av,n}=\sum^{K}_{k=1}\underbrace{ \sum^{T}_{t=1}\sum_{m:S_{m,n}=t} p_{k} q_{m}\min\left\{x_{n,k},t R_{n}\right\}}_{ \defeq \tilde{d}^{k}_{av,n}}.
\end{equation}
\begin{algorithm}[!t]
    \SetKwInOut{Input}{Input}
    \SetKwInOut{Output}{Output}
    \Input{$\mathrm{R}$,$\mathrm{C}$,$\left\{\gamma^{n}\right\}^{N}_{n=1}$}
    \Output{$\mathrm{X}^{\gamma}_{T}$}
		\For{n=1,\ldots,N}{
		    $x_{n,k} \gets 0, k\in\left\{1,\ldots,K\right\}$\;
        \While{$C_{n}>0$} {
		    $\gamma^{n}_{\acute{k},\acute{t}} \gets \max\gamma^{n}$\;
		    $x_{n,\acute{k}} \gets x_{n,\acute{k}}+\min(C_{n},R_{n})$\;
		    $\gamma^{n} \gets \gamma^{n}\setminus\left\{\gamma^{n}_{\acute{k},\acute{t}}\right\}$\;
        $C_{n} \gets C_{n}-R_{n}$\;
		                      }
											}
    \caption{Algorithm for optimal distributed caching ($\gamma$-based policy)}
\end{algorithm}
The term $\min\left\{x_{n,k},t R_{n}\right\}$ can be expanded as follows:
\begin{equation}\label{exp}
\min\left\{x_{n,k},t R_{n}\right\}=\sum^{t-1}_{i=0}\max\left\{\min\left\{x_{n,k}-iR_{n},R_{n}\right\},0\right\},
\end{equation}
and $\tilde{d}^{k}_{av,n}$ can be rewritten as:
\begin{equation}
\tilde{d}^{k}_{av,n}=\sum^{T}_{t=1}\sum_{m:S_{m,n}\geq t}p_{k}q_{m}\max\left\{\min\left\{x_{n,k}-(t-1)R_{n},R_{n}\right\},0\right\}.
\end{equation}
Equivalently,
\begin{equation}
\tilde{d}^{k}_{av,n}=\sum^{T}_{t=1}\gamma^{n}_{k,t}\max\left\{\min\left\{x_{n,k}-(t-1)R_{n},R_{n}\right\},0\right\},
\end{equation}
where $\gamma^{n}_{k,t}\defeq p_{k}P(S_{m,n}\geq t)$. We observe that for each $k$, $\tilde{d}^{k}_{av,n}$ is a monotonically increasing piecewise linear function of $x_{n,k}$, and its slope is $\gamma^{n}_{k,t}$ for $x_{n,k}\in\left((t-1)R_{n},tR_{n}\right)$. Consequently, the objective function in \text{\bf P3} is the sum of $N$ monotonically increasing piecewise linear functions, and sum of its variables are bounded by constraint (\ref{const}). Accordingly, it is maximized by maximizing the variable that corresponds to the linear function with the maximum slope.  We propose Algorithm 1 to maximize the objective function that follows a straightforward procedure using $\gamma^{n}_{k,t}$ values for each $\mathrm{SBS_{n}}$. The algorithm starts with increasing the variable $x_{n,k}$ that corresponds to the maximum slope, until the slope of $\tilde{d}^{k}_{av,n}$ changes, then it again searches for the maximum slope, and repeats this process until the sum of the variables satisfies (\ref{const}) with equality. From a computational point of view, proposed algorithm sorts the elements of set $\Gamma^{n} \defeq \left\{\gamma^{n}_{k,t}:k=1,\ldots,K;t=1,\ldots,T\right\}$ for each $n\in\mathbb{N}$. Since $| \Gamma^{n} |=KT$, the complexity of Algorithm 1 is $\mathcal{O}(NKT\log(KT))$.
The optimality of the algorithm follows from the fact that $\gamma^{n}_{k,t}\leq\gamma^{n}_{k,t-1}$ for any $(k,t)$ pair, which implies that $\tilde{d}^{k}_{av,n}$ is a concave function for each $k$. The  caching policy constructed according to Algorithm 1 is called the {\em $\gamma$-based policy}, and denoted by $\mathrm{X}^{\gamma}_{T}$. We note that when $T\leq T_{min}$, $\mathrm{X}^{\gamma}_{T}$ is the optimal policy, i.e.,  $\mathrm{X}^{\gamma}_{T}=\mathrm{X}^{\star}_{T}$. 
\vspace*{-3mm}
\subsection{Distributed Greedy Cache Allocation Scheme}
When $T>T_{min}$ it is not possible to predict the performance of $\mathrm{X}^{\gamma}_{T}$, or ensure that $d_{av}(\mathrm{X}^{\gamma}_{T},I_{T})\leq d_{av}(\mathrm{X}^{\gamma}_{T_{min}},I_{T_{min}})$.  We note that $\mathrm{X}^{\gamma}_{T_{min}}$ is the $\gamma$-based caching policy explained above for $T_{min}$. However, we know that for any $T>T_{min}$, $d_{av}(\mathrm{X}^{\gamma}_{T_{min}},I_{T})\leq d_{av}(\mathrm{X}^{\gamma}_{T_{min}},I_{T_{min}})$.
Hence, our aim is to provide a greedy distributed caching policy $\mathrm{X}^{g}_{T}$ that performs better than $\mathrm{X}^{\gamma}_{T_{min}}$, i.e., $d_{av}(\mathrm{X}^{g}_{T},I_{T}) \leq d_{av}(\mathrm{X}^{\gamma}_{T_{min}},I_{T})$. 
\begin{figure*}
    \centering
         \begin{subfigure}[b]{0.32\textwidth}
        \includegraphics[scale=0.42]{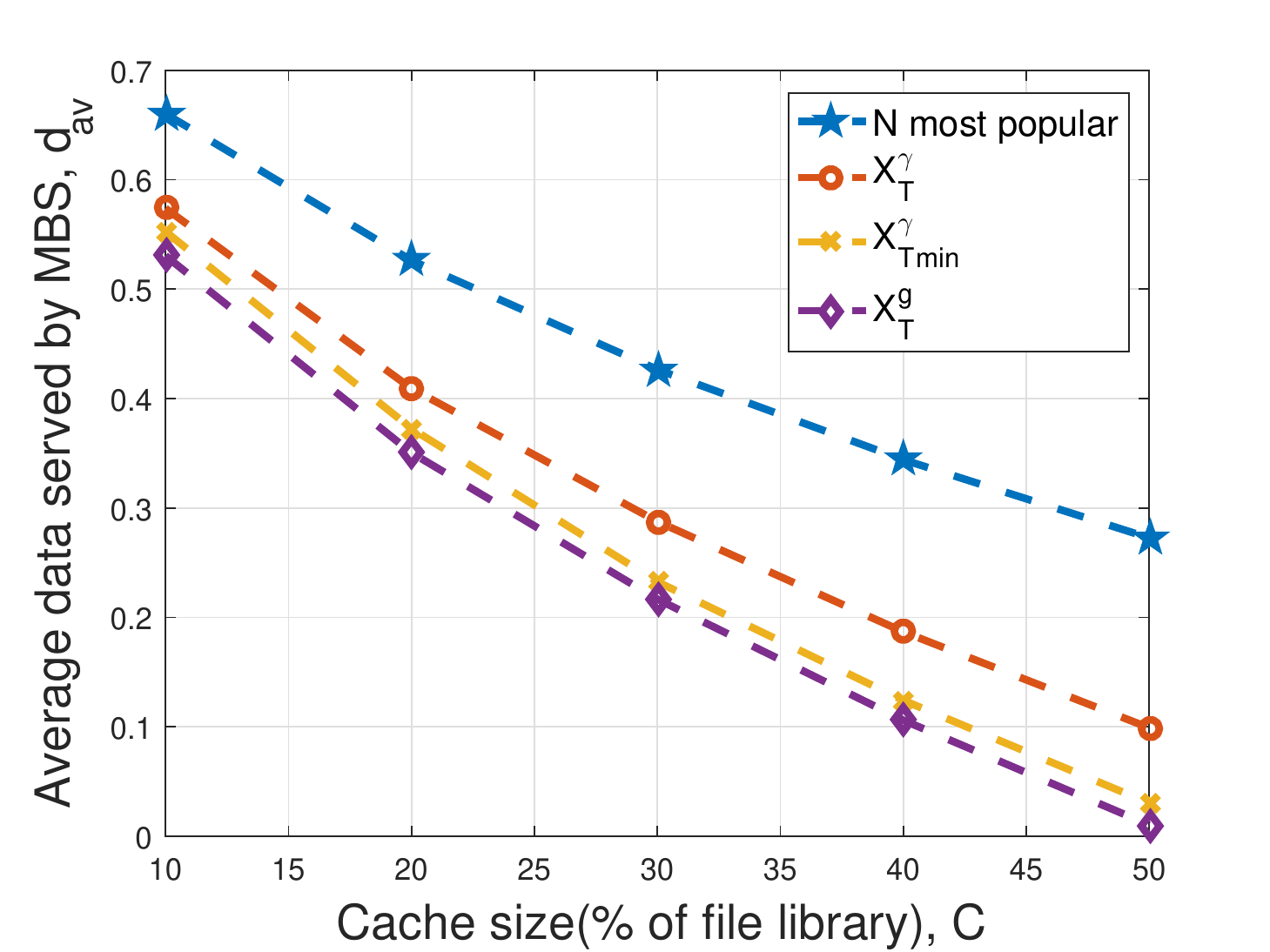}
        \caption{$T=5$ and $T_{min}=2$.}
    \end{subfigure}
    \begin{subfigure}[b]{0.32\textwidth}
        \includegraphics[scale=0.42]{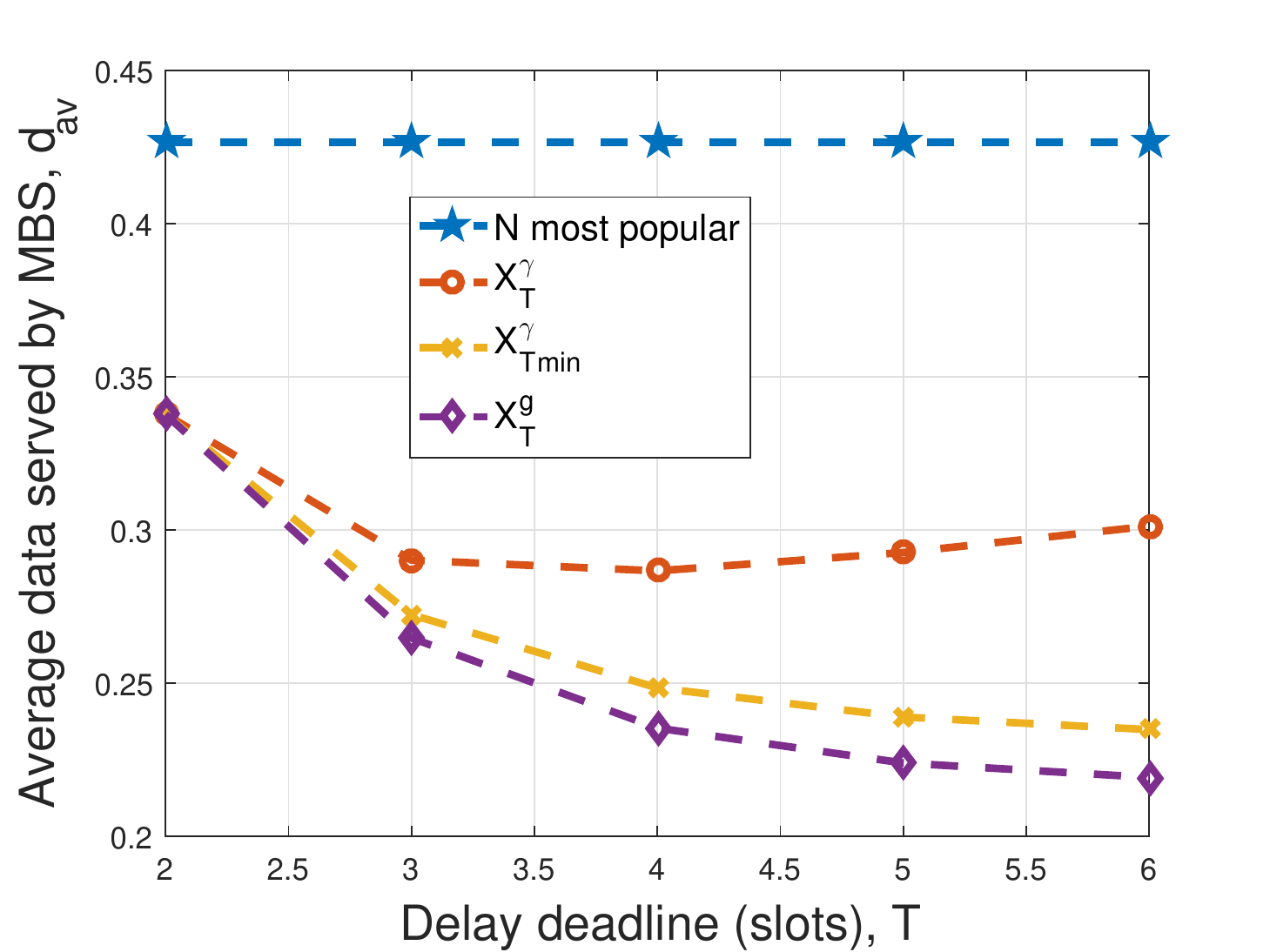}
        \caption{$C=300$ (files) and $T_{min}=2$.}
        \end{subfigure}
    \begin{subfigure}[b]{0.32\textwidth}
        \includegraphics[scale=0.42]{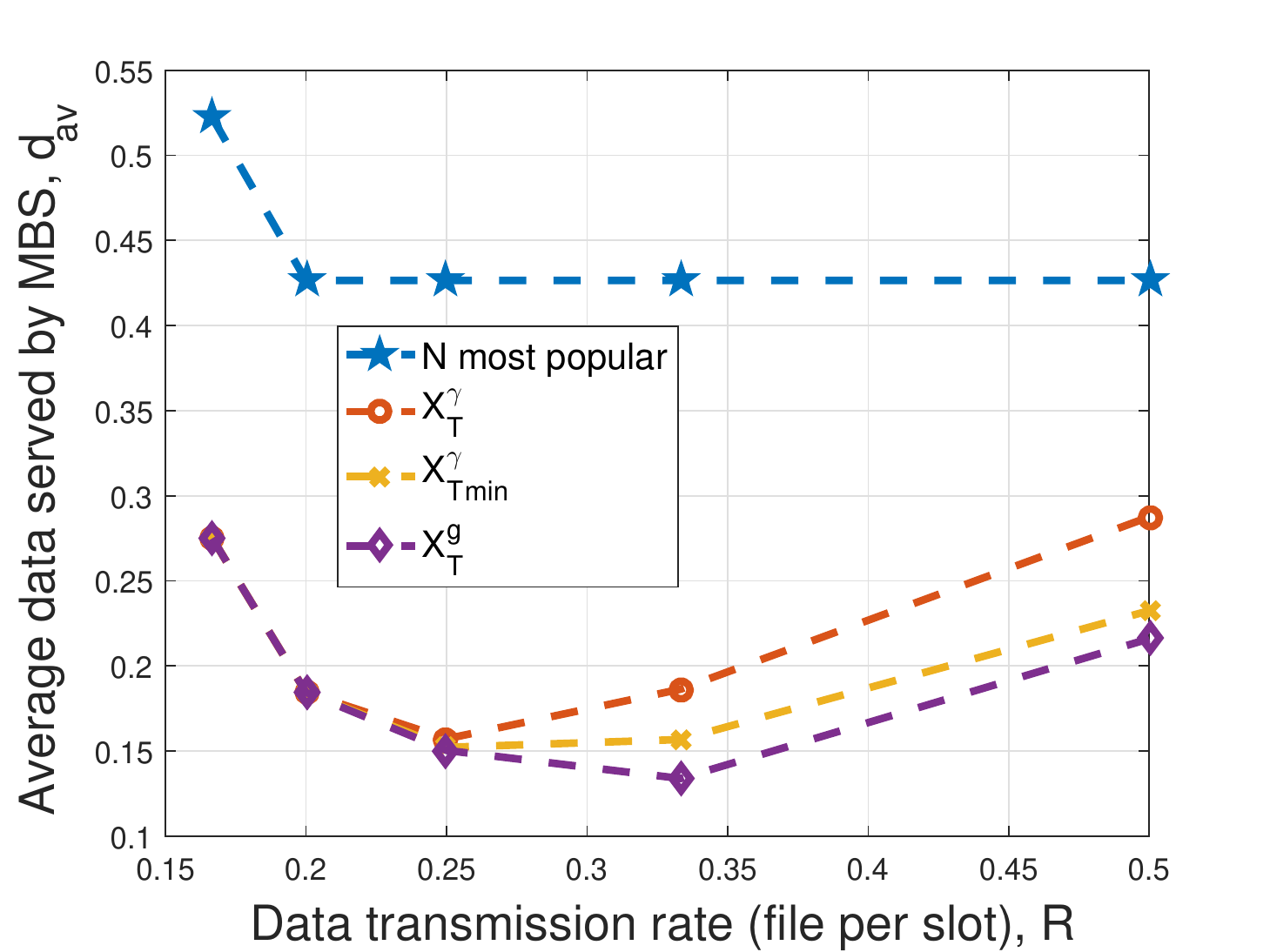}
        \caption{$C=300$ (files) and $T=5$.}
    \end{subfigure}
    \caption{The average amount of data served by the MBS (normalized by the file size)  for different values of: (a) normalized cache size $C$, (b) delay deadline $T$, and (c) data transmission rate $R=1/T_{min}$.}
\end{figure*}
\begin{algorithm}[t]
    \SetKwInOut{Input}{Input}
    \SetKwInOut{Output}{Output}
    \Input{$\mathrm{C},\mathrm{R},\mathrm{I}_{T},\mathrm{P},\mathrm{X}^{\star}_{T_{min}}$}
    \Output{$\mathrm{X}^{g}_{T}$}
		\For{n=1,\ldots,N}{
		$\Delta^{+}_{k},\Delta^{-}_{k}\gets\mathrm{NULL}: k\in\left\{1,\ldots,K\right\}$, $V^{red}\gets\left\{\right\}$\;
		$V^{+}\gets\left\{\right\}$,$V^{-}\gets\left\{\right\}$\;
		$x_{max}=max\left\{x_{n,1},\ldots,x_{n,K}\right\}$\;
    \While{$x_{max}>0$}{
		      $\acute{k}=max\left\{k:x_{n,k}\geq x_{max}\right\}$\;
		      \If{$\acute{k}\notin V^{-}$}{
		              $V^{-}\gets V^{-}\cup\left\{\acute{k}\right\}$, calculate $\Delta^{-}_{k}$ \;
																												
																			 }
					\If{$\acute{k}+1\notin V^{+}$}{																		
									$V^{+}\gets V^{+}\cup\left\{\acute{k}+1\right\}$, calculate $\Delta^{+}_{k}$\;										
																													 																	
		                                     }
            $x_{max}\gets x_{max}-R_{n}$\;
                        }
					$\Delta^{+}_{max}\gets\max\Delta^{+},\Delta^{-}_{max}\gets\max\Delta^{-}$\;				
				 \If{$\Delta^{+}_{max}>\vert\Delta^{-}_{max}\vert$}{						
											 $\acute{k}\gets\max_{k}\Delta^{+}_{k},\grave{k}\gets\max_{k}\Delta^{-}_{k}$\;
											 $\Delta^{+}_{\grave{k}},\Delta^{+}_{\acute{k}}\gets \mathrm{NULL}$\;
											 $x_{n,\acute{k}}\gets x_{n,\acute{k}}+R_{n},x_{n,\grave{k}}\gets x_{n,\grave{k}}-R_{n}$\;												
											 Go back to line 3           												
																		                       }							
											}
    \caption{Greedy algorithm for storage reallocation}
\end{algorithm} 
Our proposed method to construct $\mathrm{X}^{g}_{T}$ consists of two steps. In the first step, we obtain the optimal caching policy $\mathrm{X}^{\star}_{T_{min}}$ by executing Algorithm 1. In the second step, we follow a greedy method for cache reallocation for each cell separately, which is performed by Algorithm 2. Assume that we are reallocating the cache  for $\mathrm{SBS_{n}}$, Algorithm 2 first identifies candidate video files for cache capacity increment and reduction. In this identification process, the main criteria is the popularity of the video files, e.g., if there are several video files that have been allocated the same cache capacity, then the most popular file among those is a candidate for cache capacity increment, whereas the least popular one is a candidate for cache capacity reduction. Accordingly, let $V^{+}$ and $V^{-}$ denote the sets of indices of the video files that are candidates for cache capacity increment and reduction, respectively. After identification of a candidate file $v_{k}$, we calculate $\Delta^{-}_{k}$ if $k\in V^{-}$, or $\Delta^{+}_{k}$ if $k\in V^{+}$, whose initial values are is NULL. $\Delta^{+}_{k}$ and $\Delta^{-}_{k}$ denote the amount of change in $d_{av}$ when the cache capacity of video file $v_{k}$ is increased by $R_{n}$, or decreased by $R_{n}$ \footnote{Storage reallocation can be done with smaller sizes to improve the performance of the policy with a cost of complexity; however, due to limited space, we do not study this tradeoff  in this letter.}, respectively. In the last step, algorithm compares the values of $\Delta^{+}_{max}\defeq \max_{k}{\Delta^{+}_{k}}$ and $\Delta^{-}_{max}\defeq \max_{k}{\Delta^{-}_{k}}$. The condition $\Delta^{+}_{max}>\vert\Delta^{-}_{max}\vert$ implies that $d_{av}$ can be reduced via storage capacity reallocation. Then, the algorithm performs the following task, storage capacity of video file $v_{k}$, where $\Delta^{+}_{k}=\Delta^{+}_{max}$, is increased by $R_{n}$, and the storage capacity of video file $v_{k}$ where $\Delta^{-}_{k}=\Delta^{-}_{max}$ is decreased by $R_{n}$.

\section{Numerical Results}
For numerical simulations we consider $K=1000$ files in the library, and assume that their popularities follow a Zipf distribution with parameter $0.56$ \cite{youtube}. There are 16 SBSs located in a 2D square grid. We fix the transmission rate of each SBS to $R$, according to parameter $T_{min}$, i.e., $R=1/T_{min}$ file per slot. We consider the following Markov mobility model: a MU connected to $\mathrm{SBS_{n}}$ remains connected to the same SBS with probability $f_{n}$, or connects to one of the neighboring SBSs with equal probability. In the experiment, we consider $f_{4}=f_{13}=0.4$, $f_{7}=f_{9}=0.5$, and for the all other SBSs $f_{n}=0.3$. As a performance benchmark, we also consider the {\em $N$ most popular policy}, which simply caches the most popular $N$ files at each SBS. We remark that, when $T>T_{min}$ the value of $T$ has no impact on the performance of the $N$ most popular policy since it caches the files as a whole. On the other hand, when $T \leq T_{min}$, the value of $d_{av}$ decreases linearly with increasing $T$.

In the first experiment, we set $T_{min}=2$, $T=5$ and consider the normalized SBS cache sizes (as portion of the entire file library) $10\%,20\%,30\%,40\%,50\%$. The greedy algorithm $\mathrm{X}^{g}_{T}$ provides up to $40\%$ further reduction in the amount of data downloaded from the MBS compared to $\gamma$-based policy $\mathrm{X}^{\gamma}_{T}$ as depicted in Fig. 1(a). We also observe that the gap between the performances of $\mathrm{X}^{\gamma}_{T}$ and $\mathrm{X}^{g}_{T}$ widens with increasing cache sizes. Finally, note that the $N$ most popular policy performs quite poorly in general as it ignores the mobility patterns.

In the second experiment, we set $C=300$, $T_{min}=2$, and the delay deadline $T$ takes values from 2 to 6 time slots. Performance of the caching policies for different T values are plotted in Fig. 1(b). The key observation from the figure is that, although the average portion of the video file downloaded from the MBS  monotonically decreases with increasing $T$ under policy $\mathrm{X}^{g}_{T}$ and $\mathrm{X}^{\gamma}_{T_{min}}$, this is not the always the case for $\mathrm{X}^{\gamma}_{T}$. Note that $\mathrm{X}^{\gamma}_{T}$ mainly depends on the sojourn statistics, $P(S_{m,n}\geq t)$, over all possible paths, and when $T>T_{min}$ those statistics might be misleading because in certain paths the MU might collect all the parity bits before connecting to $\mathrm{SBS_{n}}$. In that case, storage capacity of the popular files might be increased due to sojourn statistics even though it is not required.  

In the third experiment we set $T=5$, $C=300$, and the transmission rate takes values  $1/2, 1/3, 1/4, 1/5, 1/6$ file per slot, which correspond to $T_{min}$ values of $2, 3, 4, 5, 6$ slots respectively. Although it is expected that the MBS usage decreases with the increasing transmission rate, Fig 1(c) illustrates that after a certain point the amount of data downloaded from the MBS increases with the transmission rate under all policies. This is because,  $T_{min}$ decreases when the rate increases and the difference between $T_{min}$ and $T$  widens, as a result of which the performance becomes worse. 

\section{Conclusions}
In this letter, we studied mobility and popularity aware content caching for a heterogeneous network with MDS-coded caching at the SBSs. Assuming a maximum download time requirement $T$, for each request, we first defined the threshold $T_{min}$ on $T$, below which some bits of the request must be downloaded from the MBS. Then, we obtained the optimal distributed caching policy when $T\leq T_{min}$,  called the $\gamma$-based policy, which minimizes the amount of data that need to be downloaded from the MBS. Then, we utilized the parameter $T_{min}$ and the $\gamma$-based policy for $T=T_{min}$ to obtain a greedy caching policy for $T > T_{min}$. Consequently, we showed how to design a coded caching policy according to $T_{min}$ and performed various simulations to demonstrate that the utilization of $T_{min}$ improves the performance significantly.

\ifCLASSOPTIONcaptionsoff
  \newpage
\fi



%
\bibliographystyle{IEEEtran}
\bibliography{IEEEabrv,comlet.bib}

%





\end{document}